\documentstyle[sprocl,epsfig]{article}

\def\be{\begin{equation}}
\def\ee{\end{equation}}
\def\bea{\begin{eqnarray}}
\def\eea{\end{eqnarray}}

\newcommand{\newc}{\newcommand} 
\newc{\lra}{\leftrightarrow} 
\newc{\beq}{\begin{equation}} 
\newc{\eeq}{\end{equation}} 
\newc{\barr}{\begin{eqnarray}} 
\newc{\earr}{\end{eqnarray}} 
\begin{document} 
\title{SEARCHING FOR SUPERSYMMETRIC DARK MATTER. THE MODULATION
EFFECT DUE  TO THE CAUSTIC RINGS . }

\author{J. D. VERGADOS} 

\address{Theoretical Physics Section, University of Ioannina, GR-45110, 
Greece\\E-mail:Vergados@cc.uoi.gr} 

\maketitle\abstracts{  The detection of the theoretically expected dark matter
is central to particle physics and cosmology. Current fashionable supersymmetric
models provide a natural dark matter candidate which is the lightest
supersymmetric particle (LSP). 
The theoretically obtained event rates
are usually  very low or even undetectable.
 So the experimentalists would like to exploit the modulation effect. 
In the
present paper we study a specific class of non-isothermal models
involving flows of caustic rings.
We find that the modulation effect arising from such models is smaller
than that predicted by the isothermal models.
}
\section{Introduction}
In recent years the consideration of exotic dark matter has become necessary
in order to close the Universe  \cite{Jungm}.  Recent data from the 
Supernova Cosmology Project  ~\cite {SPF}
$^,$~\cite {SCP} suggest that the situation can be adequately
described by  a barionic component $\Omega_B=0.1$ along with the exotic 
components $\Omega _{CDM}= 0.3$ and $\Omega _{\Lambda}= 0.6$ (see also
Turner, these proceedings). 

 Since this particle is expected to be very massive, $m_{\chi} \geq 30 GeV$, and
extremely non relativistic with average kinetic energy $T \leq 100 KeV$,
it can be directly detected ~\cite{JDV}$^-$\cite{KVprd} mainly via the
recoiling nucleus.

 Using an effective supersymmetric Lagrangian at the  
quark level, see e.g. Jungman et al 
~\cite{Jungm} and references therein , a quark model for the nucleon
 ~\cite{Dree,Chen}
and nuclear  wave functions ~\cite{KVprd} one can obtain the needed
detection rates.  They are typically very low. So experimentally one 
would like to exploit the modulation of the event rates due to the 
earth's revolution around the sun. In our previous work
\cite{JDV99}$^,$ \cite{JDV99b} we found
anhanced modulation, if one uses appropriate asymmetric velocity
distribution.  
 The isolated galaxies are, however, surrounded by cold dark matter
, which,  due to gravity, keeps falling continuously on 
them from all directions \cite{SIKIVIE}.
It is the purpose of our present paper to exploit the results of
such a scenario.

\section{The Basic Ingredients for LSP Nucleus Scattering}

The differential cross section can be cast in the form 
\cite{JDV99b}:
\beq
d\sigma (u,\upsilon) = \frac{du}{2 (\mu _r b\upsilon )^2} [(\bar{\Sigma} _{S} 
                   +\bar{\Sigma} _{V}~ \frac{\upsilon^2}{c^2})~F^2(u)
                       +\bar{\Sigma} _{spin} F_{11}(u)]
\label{2.9}
\eeq
with 
\beq 
\bar{\Sigma} _{S} = \sigma_0 (\frac{\mu_r}{m_N})^2  \,
 \{ A^2 \, [ (f^0_S - f^1_S \frac{A-2 Z}{A})^2 \, ] 
\label{2.10}
\eeq 
 The functions $\bar{\Sigma}_{spin}$, associated with the spin, and
the small coherent term, $\bar{\Sigma}_V$, associated with the vector
contribution, are not going to be discussed further (see oue earlier
work \cite{JDV99b}).
In the above expression $m_N$ is the proton mass,
$\mu_r$ is the reduced mass, $F(u)$ and $F_{11}(u)$ are  the usual and
isovector spin nuclear form factors and
 $ u = q^2b^2/2$,
with b the harmonic oscillator size parameter and q the momentum transfer
to the nucleus.
The scale is set by$  
\sigma_0 = \frac{1}{2\pi} (G_F m_N)^2 \simeq 0.77 \times 10^{-38}cm^2$
The quantity u is also related to the experimentally measurable energy transfer
Q via the relations $Q=Q_{0}u,~ Q_{0} = [A m_{N} b^2]^{-1}$ 
The needed parameters from SUSY are $f^0_S, f^1_S$ (isoscalar and 
isovector sacalar respectively).
 The differential detection rate for a particle with 
velocity ${\boldmath \upsilon}$ 
and a target with mass $m$ is 
\beq
dR = \frac{\rho (0)}{m_{\chi}} \frac{m}{A m_N} 
d\sigma (u,\upsilon)
|{\boldmath \upsilon}| 
\label{2.18}  
\eeq
where $d\sigma(u,\upsilon )$ is given by Eq. ( \ref{2.9})
 One normally assumes $\rho (0) = 0.3 GeV/cm^3$ as the LSP density 
in our vicinity. 
\section{Convolution of the Event Rate}
 
In this section we will examine the consequences of the earth's
revolution around the sun (the effect of its rotation around its axis is
expected to be negligible) i.e. the modulation effect. 

 Following Sikivie we will consider $2 \times N$ caustic rings, (i,n)
, i=(+.-) and n=1,2,...N (N=20 in the model of Sikivie et al),
each of which
contributes to the local density a fraction $\bar{\rho}_n$ of the
the summed density $\bar{\rho}$ of each type $i=+,-$. and has
velocity ${\bf y}_n=(y_{nx},y_{ny},y_{nz})$
, in units of $\upsilon_0=220~Km/s$, with respect to the
galactic center.

We find it convenient to choose the z-axis 
in the direction of the motion of the
the sun, the y-axis is normal to the plane of the galaxy and 
the x-axis is in the radial direction. 
The needed quantities are takan from the 
work of Sikivie (table 1 of last Ref. \cite{SIKIVIE}) by the 
definitions
$y_n=\upsilon_n/\upsilon_0
,y_{nz}=\upsilon_{n\phi}/\upsilon_0
,y_{nx}=\upsilon_{nr}/\upsilon_0
,y_{ny}=\upsilon_{nz}/\upsilon_0$
. This leads to a
velocity distribution of the form:
\beq
f(\upsilon^{\prime}) = \sum_{n=1}^N~\delta({\bf \upsilon} ^{'}
    -\upsilon_0~{\bf y}_n)
\label{3.1}  
\eeq
The velocity of the earth around the
sun is given by \cite{KVprd}. 
\beq
{\bf \upsilon}_E \, = \, {\bf \upsilon}_0 \, + \, {\bf \upsilon}_1 \, 
= \, {\bf \upsilon}_0 + \upsilon_1(\, sin{\alpha} \, {\bf \hat x}
-cos {\alpha} \, cos{\gamma} \, {\bf \hat y}
+cos {\alpha} \, sin{\gamma} \, {\bf \hat z} \,)
\label{3.6}  
\eeq
where $\alpha$ is the phase of the earth's orbital motion, $\alpha =0$
around second of June. In the laboratory frame we have
\cite{JDV99b}
$ {\bf \upsilon}={\bf \upsilon}^{'} \, - \, {\bf \upsilon}_E \,$ 


\section{Undirectional Event Rates}

Integrating Eq. (\ref {2.18}) we obtain for the total
undirectional rate
\beq
R =  \bar{R}\, t \, \frac{2 \bar{\rho}}{\rho(0)}
          [1 - h(a,Q_{min})cos{\alpha}] 
\label{3.55}  
\eeq
where $Q_{min}$ is the energy transfer cutoff imposed by the detector.
and $a =[\sqrt{2} \mu _rb\upsilon _0]^{-1}.  $ Also
$\rho_{n}=d_n/\bar{\rho}
,\bar{\rho}=\sum_{n=1}^N~d_n$ (for each flow +,-).
In the Sikivie model
\cite {SIKIVIE} $2\bar{\rho}/\rho(0)=1.25$. 
In the above expressions $\bar{R}$ is the rate obtained 
\cite {JDV} by neglecting the folding with the LSP velocity and the
momentum transfer dependence, i.e. by
\beq
\bar{R} =\frac{\rho (0)}{m_{\chi}} \frac{m}{Am_N} \sqrt{\langle
v^2\rangle } [\bar{\Sigma}_{S}+ \bar{\Sigma} _{spin} + 
\frac{\langle \upsilon ^2 \rangle}{c^2} \bar{\Sigma} _{V}]
\label{3.39b}  
\eeq
and it contains all SUSY parameters except $m_{chi}$
 The modulation is described in terms of the parameter $h$. 
 The effect of folding
with LSP velocity on the total rate is taken into account via the quantity
$t$. 

We like to
stress that it is a common practice to extract the LSP nucleon cross 
section from the  the meusured event rates in order to 
compare with the SUSY predictions . In
such analyses the factor of $ t$ is commonly ommitted. It is clear, 
however, that,
in going from the data to the cross section, one should
divide by $t$. 

The undirectional differential rate takes the form
\beq
\langle \frac{dR}{du} \rangle  = \bar{R} \frac{2 \bar{\rho}}{\rho(0)}
                   t T(u) [1 - \cos \alpha~ H(u)]
\label{3.31}  
\eeq

The factor 
$T(u)$ takes care of the u-dependence of the unmodulated differential rate. It
is defined so that
\beq
 \int_{u_{min}}^{u_{max}} du T(u)=1.
\label{3.30a}  
\eeq
i.e. it is the relative differential rate. $u_{min}$ is determined by the energy 
cutoff due to the performance of the detector. $u_{max}$ is determined by the 
escape velocity $\upsilon_{esc}$ via the relations:
$u_{max}=max(\frac{y_{n} ^2}{a^2}~~ n=1,2,...,N$.
On the other hand
$H(u)$ gives the energy transfer dependent
modulation amplitude (relative to the unmodualated one).
\section{Discussion of the Results and Conclusions}

We have calculated the the total event rates 
for elastic LSP-nucleus scattering 
including realistic nuclear form factors. We focused our attention on those aspects of the problem, which do not depend on the parameters of supersymmetry
other than the LSP mass.
 The parameter $\bar {R}$, normally calculated in SUSY theories, was not
considered in this work. The interested reader is referred to the 
literature
, for a review \cite {Jungm} and references therin and, in our notation, 
to our previous work \cite {JDV} $^,$ \cite {KVprd}.

 We specialized our results for the target $^{127}I$.  We considered
the effects of the detector energy cutoff, by considering  two  typical
cases $Q_{min}=10$ and $Q_{min}=20$ KeV. 
We assumed that the LSP density in our vicinity and the
velocity spectrum is that of caustic rings of Sikivie et al 
\cite{SIKIVIE}. 
\setlength{\unitlength}{1mm}
\begin{figure}
\vspace*{1.8cm}
\begin{picture}(190,30)
\put(10,0){\epsfxsize=5cm \epsfbox{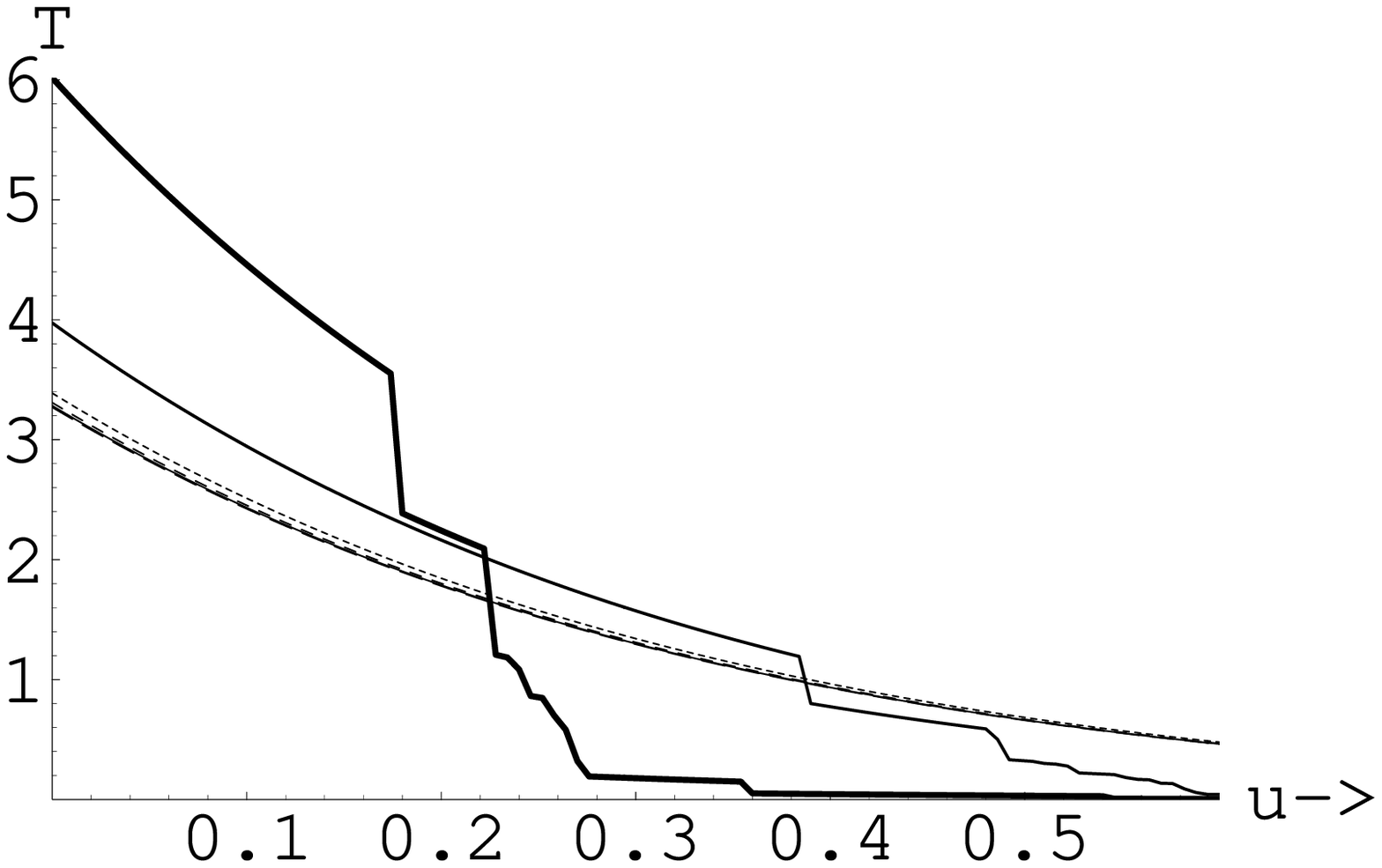}}
\put(70,0){\epsfxsize=5cm \epsfbox{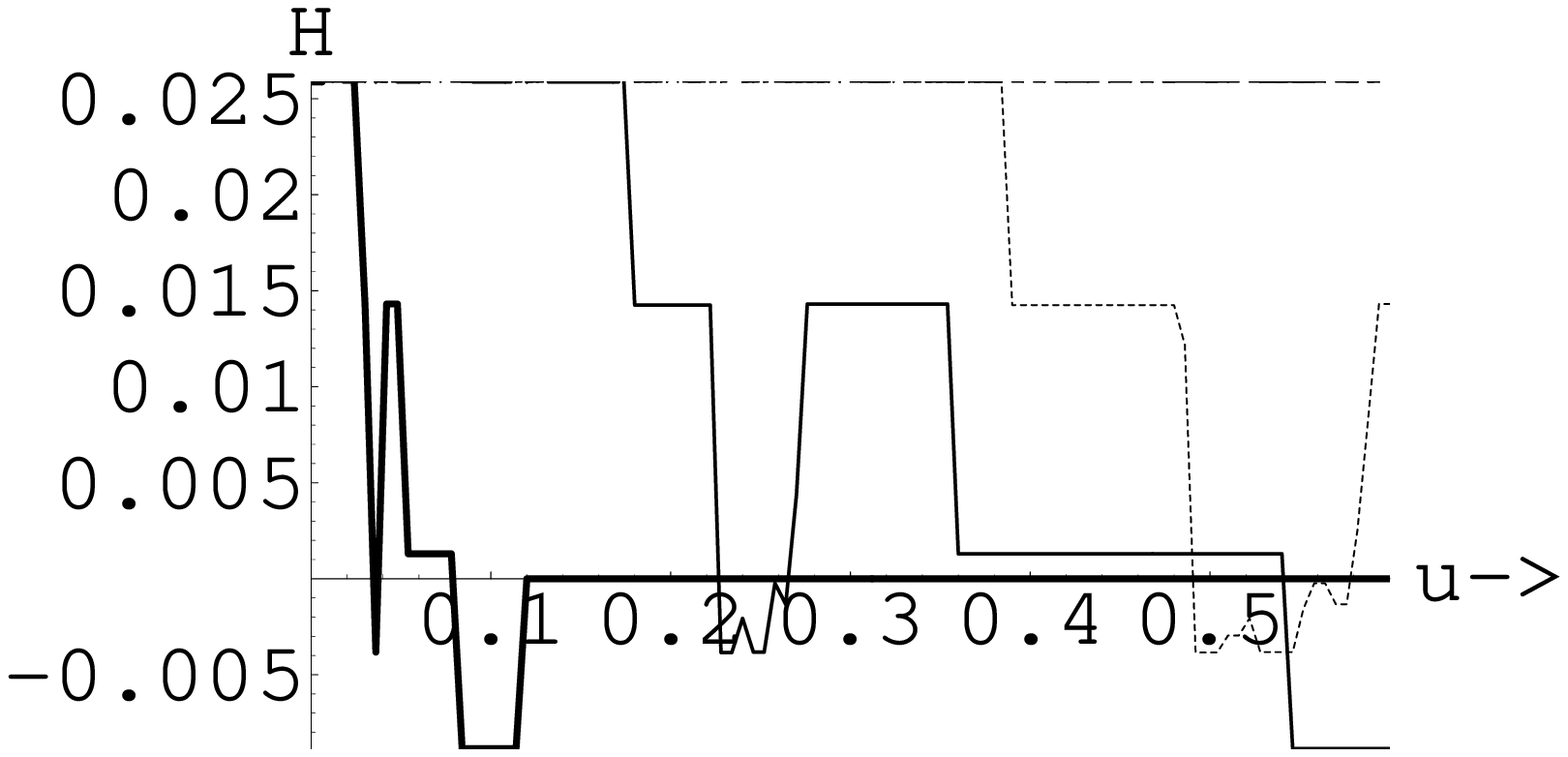}}
\end{picture}
\vspace*{-2.0cm}
\caption[]{The quantities $T(u)$ and $H(u)$ entering the differential
amplitude. Thick solid line corresponds to $m_{chi}=30~GeV$ the intermediate
thickness line to $m_{chi}=80~ GeV$, the fine line to $m_{chi}=100~GeV$.
The rest correspond to larger LSP masses and fall on top of each other.f
}
\label{fig.1}
\end{figure}

The total rates are 
described in terms of  $t$ and $h$.
In TABLE I we show how they vary the detector energy cutoff
and the LSP mass.

The parameters $T(u)$ and $H(u)$ entering the differential amplitude
are shown in Fig. 1. The shape of $T(u)$ is analogous to that of the
isothermal models except that the maximum occurs at $u=0.0$, rather than
at $u=0.1$. The function $H(u)$ shows 
oscillations, which result in a smaller total modulation.
Another way of understanding how the cancellations arize is to note 
that for some rings $y_{nz}>1$, while for  others $y{_nz}<1$.
\begin{table}[t]  
\caption{The quantities $t$ and $h$ in the case of the
target $_{53}I^{127}$ for various LSP masses and $Q_{min}$ in KeV 
(for definitions see text). 
}
\begin{center}
\footnotesize
\begin{tabular}{|l|c|rrrrrrr|}
\hline
\hline
& & & & & & & &     \\
&  & \multicolumn{7}{|c|}{LSP \hspace {.2cm} mass \hspace {.2cm} in GeV}  \\ 
\hline 
& & & & & & & &     \\
Quantity &  $Q_{min}$  & 10  & 30  & 50  & 80 & 100 & 125 & 250   \\
\hline 
& & & & & & & &     \\
t    &0.0&1.451& 1.072& 0.751& 0.477& 0.379& 0.303& 0.173\\
h    &0.0&0.022& 0.023& 0.024& 0.025& 0.026& 0.026& 0.026\\
\hline 
& & & & & & & &     \\
t    &10.0&0.000& 0.226& 0.356& 0.265& 0.224& 0.172& 0.098\\
h    &10.0&0.000& 0.013& 0.023& 0.025& 0.025& 0.026& 0.026\\
\hline 
& & & & & & & &     \\
t    &20.0&0.000& 0.013& 0.126& 0.139& 0.116& 0.095& 0.054\\
h    &20.0&0.000& 0.005& 0.017& 0.024& 0.025& 0.026& 0.026\\
\hline
\hline
\end{tabular}
\end{center}
\end{table}

The maximum occurs around the 2nd of December, something already noticed
by Sikivie et al \cite{SIKIVIE}. Furthermore the modulation
is small, $h=0.25$, i.e. a $5\%$ difference betwen the maximum and the 
minimum (see TABLE I).
It is a bit smaller than that of the symmetric models, but
a lot smaller than that predicted by the  asymmetric
ones \cite{JDV99,JDV99b}, i.e $h=0.46$

\section*{Acknowledgments}

The author would like to 
acknowledge partial support of this work by 
TMR No  ERB FMAX-CT96-0090.   

\end{document}